\documentclass[aps,showpacs,showkeys,amsmath,amssymb,twocolumn,floatfix,superscriptaddress]{revtex4}
\usepackage{graphicx}
\usepackage{amssymb}
\usepackage{dcolumn}
\usepackage{bm}
\usepackage{color}
\begin{document}

\title{Lattice QCD-based equations of state at vanishing net-baryon density} 

\author{M.~Bluhm}
\email{mbluhm@to.infn.it}
\affiliation{Dipartimento di Fisica, Universit\`{a} degli Studi di Torino \& INFN, Sezione di Torino, via Giuria 1, I-10125 Torino, Italy}
\author{P.~Alba}
\affiliation{Dipartimento di Fisica, Universit\`{a} degli Studi di Torino \& INFN, Sezione di Torino, via Giuria 1, I-10125 Torino, Italy}
\author{W.~Alberico}
\affiliation{Dipartimento di Fisica, Universit\`{a} degli Studi di Torino \& INFN, Sezione di Torino, via Giuria 1, I-10125 Torino, Italy}
\author{A.~Beraudo}
\affiliation{Physics Department, Theory Unit, CERN, CH-1211 Gen\`{e}ve 23, Switzerland}
\author{C.~Ratti}
\affiliation{Dipartimento di Fisica, Universit\`{a} degli Studi di Torino \& INFN, Sezione di Torino, via Giuria 1, I-10125 Torino, Italy}



\begin{abstract}
We present realistic equations of state for QCD matter at vanishing net-baryon density which embed recent lattice QCD results at high temperatures combined with a hadron resonance gas model in the low-temperature, confined phase. In the latter, we allow an implementation of partial chemical equilibrium, in which particle ratios are fixed at the chemical freeze-out, so that a description closer to the experimental situation is possible. Given the present uncertainty in the determination of the chemical freeze-out temperature from first-principle lattice QCD calculations, we consider different values within the expected range. The corresponding equations of state can be applied in the hydrodynamic modeling of relativistic heavy-ion collisions at the LHC and at the highest RHIC beam energies. Suitable parametrizations of our results as functions of the energy density are also provided.  
\end{abstract}

\maketitle

\section{Introduction \label{sec:1}}

In the relativistic heavy-ion collisions at RHIC (Relativistic Heavy-Ion Collider) and LHC (Large Hadron Collider), a hot deconfined state of strongly interacting matter is transiently created, the Quark-Gluon Plasma (QGP). This form of QCD matter is believed to have existed in the very first moments of our universe. As the produced hot and dense system cools down during its expansion, matter undergoes a transition from the QGP phase into a state dominated by color-confined, massive hadronic degrees of freedom. The nature of this phase transformation has been determined at vanishing baryon-chemical potential by first-principle lattice QCD simulations: it is an analytic crossover, taking place over a broad region of temperatures $T$~\cite{Aoki06}. The value of the (pseudo-) critical temperature $T_c$ associated with this confinement transition depends to some extent on the considered order-parameter. 
For example, the {\it Wuppertal-Budapest} (WB) and {\it hotQCD} collaborations found comparable values for chiral symmetry restoration: $T_c=$($155\pm 6$)~MeV in~\cite{Borsanyi09} and $T_c=$($154\pm 9$)~MeV in~\cite{Bazavov12}, respectively. 

The collective flow dynamics of the bulk of matter created in heavy ion collisions can be successfully modeled by means of relativistic hydrodynamics (cf.~e.g.~the reviews in~\cite{Kolb03,Gale13}), starting from a stage immediately after thermalization until the kinetic freeze-out of final state hadrons. Assuming local thermal equilibrium, the conservation equations for energy, momentum and for the additionally conserved charges (net-baryon number $N_B$, net-electric charge $N_Q$ and net-strangeness $N_S$) drive the evolution of the system. An essential ingredient for this modeling is the equation of state (EoS), which provides locally a relation between energy density $\epsilon$, pressure $p$ and the densities $n_B$, $n_Q$ and $n_S$ of the conserved charges. The parameter controlling the acceleration of the fluid collective flow due to pressure gradients is the speed of sound, $c_s=\sqrt{\partial p/\partial\epsilon}$. 

A quantitative comparison of hydrodynamic simulations with the observed collective flow behavior revealed that the evolution of the system can be described by nearly ideal hydrodynamics, cf.~e.g.~\cite{Romatschke07,Luzum08,Song09,Song11,Schenke11,Song13,Luzum13}. In these studies, a uniquely small ratio of shear viscosity $\eta$ to entropy density $s$ of the hot matter was determined, cf.~also the reviews in~\cite{Schafer09,Teaney10}. This led to our current understanding of the QGP as a strongly coupled, nearly perfect fluid~\cite{Gyulassy05,Shuryak05,Heinz05}. Assuming the conservation of entropy, i.e.~neglecting the viscous entropy production associated with such a small $\eta/s$~\cite{Shen10}, one needs to know the EoS only along adiabatic paths. In this work, we concentrate on the situation of a vanishing $n_B$, i.e.~we consider the path $n_B/s=0$. We note that in the thermal system created in a heavy-ion collision one always has $n_S=0$, while $n_Q$ (in the case of a partial stopping at the lowest 
center-of-mass energies) is related to $n_B$. 

A rigorous determination of the equation of state in thermal and chemical equilibrium for $n_B=0$ in the non-perturbative regime of QCD can be achieved with lattice gauge theory simulations. These reach nowadays unprecedented levels of accuracy. A basic quantity for the EoS is the interaction measure $I=\epsilon-3p$, which has been calculated in~\cite{Bazavov09,Cheng10} and in~\cite{Aoki05,Borsanyi10,Borsanyi11}. The numerical results for $I(T)/T^4$ in~\cite{Bazavov09,Cheng10} show significant differences from those in~\cite{Borsanyi10,Borsanyi11} in the transition region. In this work, we opt for utilizing the recent, continuum-extrapolated lattice QCD data from the WB-collaboration presented in~\cite{Borsanyi11}, corresponding to a system of 2+1 quark flavors with physical quark masses. 
By combining a suitable parametrization of these lattice QCD results with a hadron resonance gas (HRG) model in thermal and chemical equilibrium, we construct a baseline QCD equation of state for $n_B=0$. 

The focus of our work lies, however, on the implementation of partial chemical equilibrium, i.e.~a non-equilibrium situation, in the hadronic phase. In this way, one can properly account for the actual chemical composition in the confined phase, an issue which is not addressed within equilibrium lattice QCD thermodynamics. This is known to be of importance in order to reproduce not only the experimentally observed flow and $p_T$-spectra, but also the correct particle ratios~\cite{Hirano02}. 

Because of the present uncertainty in the exact value of the chemical freeze-out temperature $T_{ch}$, cf.~\cite{Borsanyi13}, we consider various values for $T_{ch}$ within the expected range, below which the HRG is assumed to be in partial chemical equilibrium. In this way, 
different realistic QCD equations of state are obtained, which can be used in the hydrodynamic simulations of relativistic heavy-ion collisions for LHC and RHIC top beam energies at mid-rapidity when net-baryon density effects can be neglected. Having such QCD equations of state at hand will allow to study the possible impact of a variation of $15$~MeV in $T_{ch}$ on particle spectra as well as a 
more controlled determination of the QGP transport properties, as for example the shear and bulk viscosity coefficients. At smaller beam energies, effects of a non-vanishing net-baryon density become important. Corresponding QCD equations of state will be presented in a forthcoming publication. 

The equation of state of QCD matter has been the subject of numerous studies in the literature. Among different other approaches, we mention combinations of the HRG model with an effective theory of QCD~\cite{Laine06}, with a phenomenological model for QCD thermodynamics~\cite{Bluhm07} and with various parametrizations~\cite{Chojnacki07,Chojnacki08,Song08,Huovinen09} of lattice QCD results. Developments in using a parametrization of lattice QCD results for finite $n_B$ were recently reported in~\cite{Huovinen11,Huovinen12}. Moreover, in~\cite{Steinheimer10} an EoS, describing both the QGP and the hadronic phase based on one effective model approach, was constructed and applied in finite-$n_B$ hydrodynamics studies (see also further developments in~\cite{Steinheimer11}). 

The paper is organized as follows: in section~\ref{sec:2}, we discuss briefly the employed lattice QCD results~\cite{Borsanyi11} and their combination with a HRG model in thermal and chemical equilibrium. Section~\ref{sec:3} deals with the inclusion of partial chemical equilibrium in the description of the hadronic phase. In section~\ref{sec:4}, we discuss the obtained QCD equations of state and provide practical parametrizations of our results. 

\section{Construction of a Lattice QCD-based EoS\label{sec:2}}

\begin{figure*}[t]
\centering
\vspace{5mm}
\includegraphics[width=0.45\textwidth]{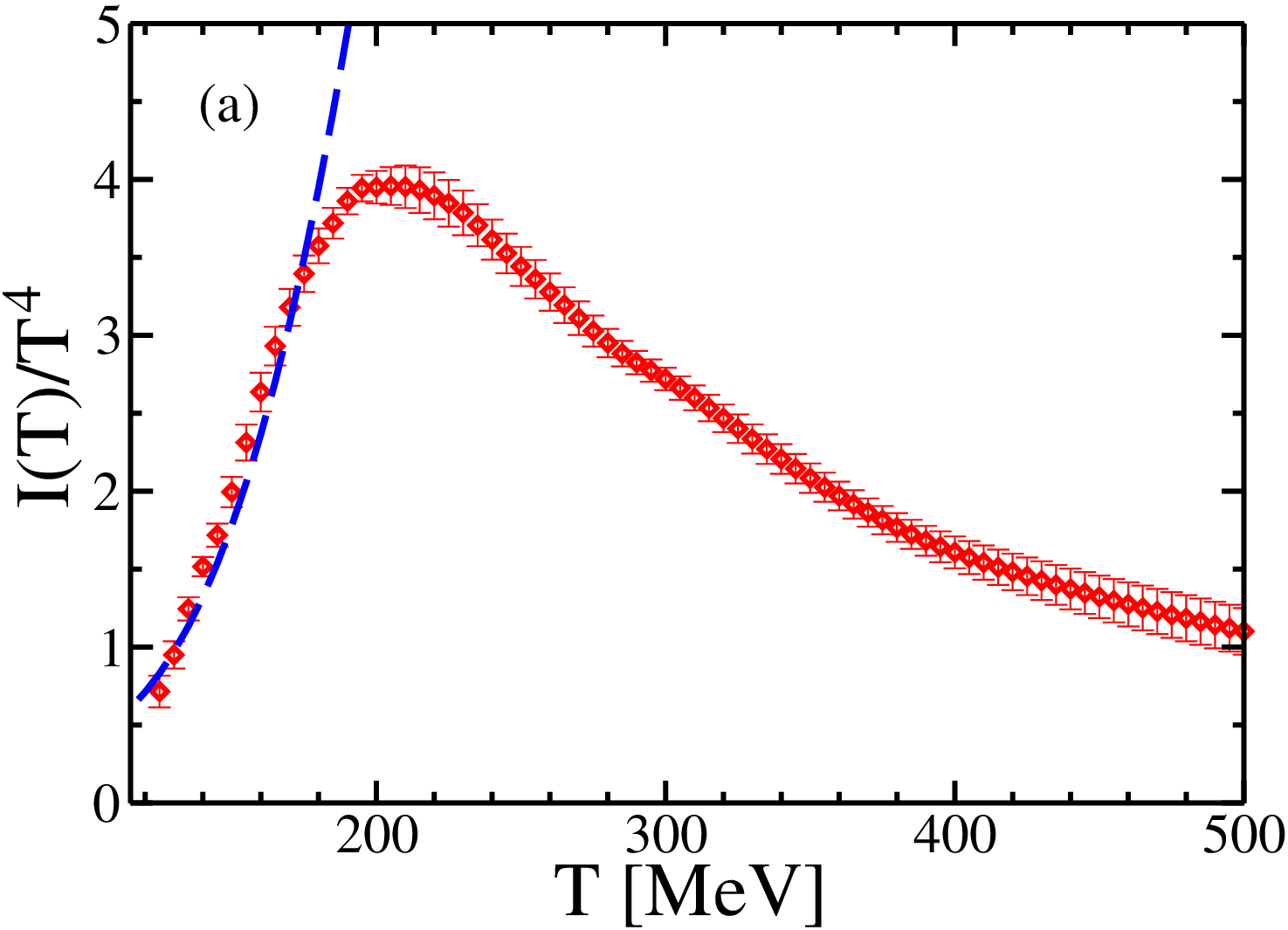}
\hspace{3mm}
\includegraphics[width=0.45\textwidth]{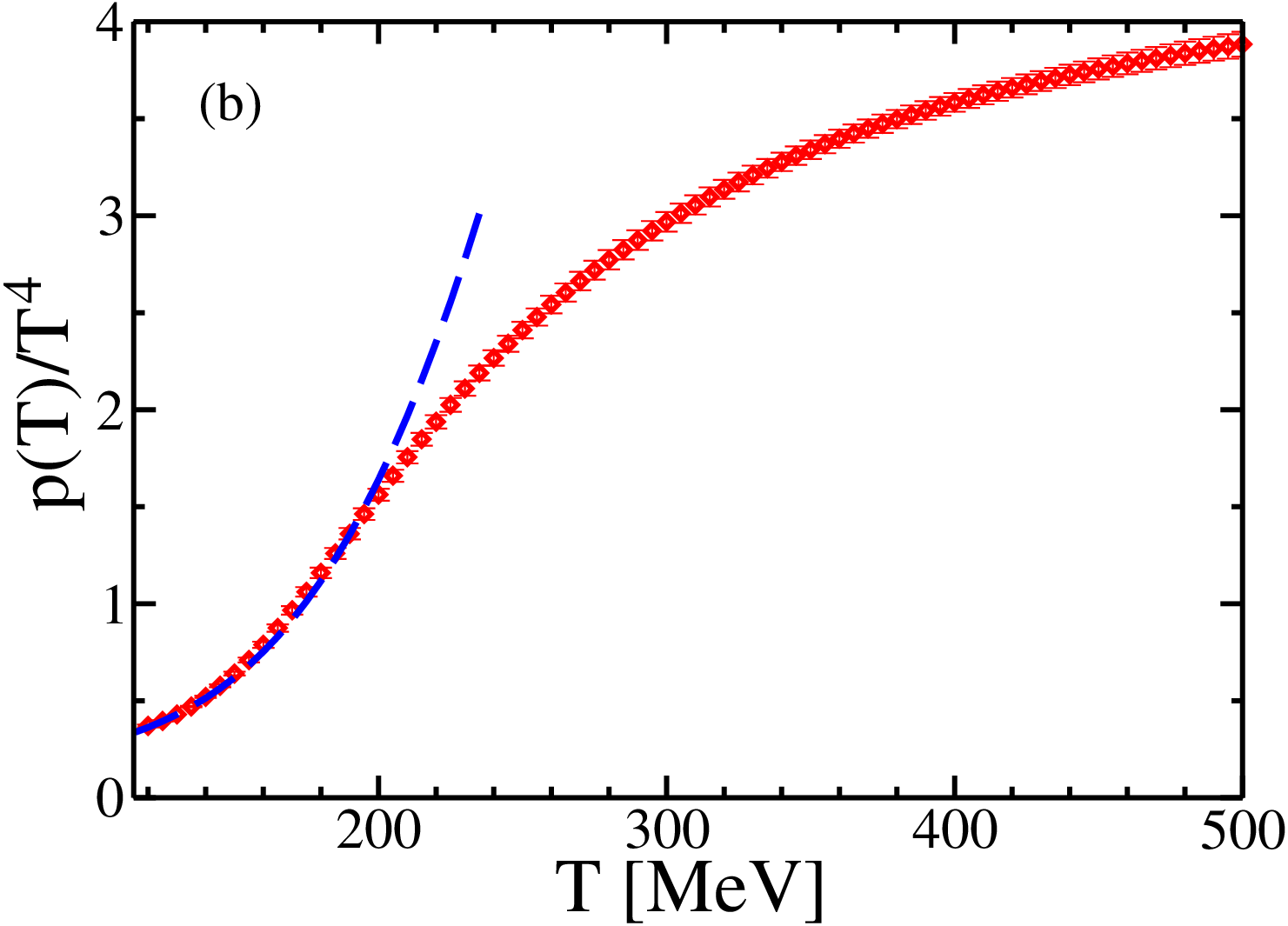} 
\caption[]{\label{Fig:1} (Color online) Scaled interaction measure $I(T)/T^4$ (panel (a)) and pressure $p(T)/T^4$ (panel (b)) as functions of the temperature $T$. The symbols depict the lattice QCD results from~\cite{Borsanyi11}, while the dashed curves show the results of the employed HRG model in thermal and chemical equilibrium.}
\end{figure*} 
In~\cite{Borsanyi11}, continuum-extrapolated lattice gauge theory results of QCD thermodynamics for 2+1 quark flavors with physical mass parameters were presented. The corresponding results for the scaled interaction measure $I(T)/T^4$ and for the scaled pressure $p(T)/T^4$ are depicted in Fig.~\ref{Fig:1} panels (a) and (b), respectively. A suitable parametrization of the results for $I(T)/T^4$ as a function of $T$, which provides an accurate description within the given error-band, can be found by employing a similar fit function as the one used in~\cite{Borsanyi10}. From this, the other thermodynamic quantities follow via thermodynamic identities: the scaled pressure is determined by a definite $T$-integral of $I(T)/T^5$, while $\epsilon(T)=3p(T)+I(T)$ and $s(T)=(\epsilon(T)+p(T))/T$. This yields for $s(T)$ extrapolated to $T=800$~MeV a value of about $82.5\%$ of the {\it Stefan-Boltzmann} limit for a non-interacting gas of 3 massless quark flavors. 

The thermodynamics of QCD matter in the hadronic phase can be well accounted for by the HRG model describing hadronic matter in thermal and chemical equilibrium, cf.~e.g.~\cite{Karsch03,Tawfik05}. The pressure of the model in the thermodynamic limit is given by 
\begin{eqnarray}
\nonumber
 p(T,\{\mu_k\}) & = & \sum_k (-1)^{B_k+1} \frac{d_kT}{(2\pi)^3} \int d^3\vec{p} \,\,\ln\Big[1 +\\
\label{equ:pressureHRG}
 & & \,\,\, (-1)^{B_k+1} e^{-(\sqrt{\vec{p}^{\,2}+m_k^2}-\mu_k)/T}\Big] \,,
\end{eqnarray}
where the sum is taken over all hadronic (including resonances) states $k$ (baryons and anti-baryons being summed independently) included in the model. In Eq.~(\ref{equ:pressureHRG}), $d_k$ and $m_k$ denote the degeneracy factor and the mass, and $\mu_k$ is the chemical potential of the hadron-species $k$. In chemical equilibrium, the latter reads $\mu_k=B_k\mu_B+Q_k\mu_Q+S_k\mu_S$, where $B_k$, $Q_k$ and $S_k$ are the respective quantum numbers of baryon charge, electric charge and strangeness, while $\mu_B$, $\mu_Q$ and $\mu_S$ denote the chemical potentials associated with $n_B$, $n_Q$ and $n_S$. 

Other thermodynamic quantities follow from standard relations, e.g.~$s\!=\!(\partial p/\partial T)_{\mu_k}$. The particle number density of species $k$, $n_k\!=\!(\partial p/\partial\mu_k)_T$, is given by the momentum-integral 
\begin{equation}
\label{equ:particledensity}
 n_k(T,\mu_k)=\frac{d_k}{(2\pi)^3}\int d^3\vec{p} \frac{1}{(-1)^{B_k+1}+e^{(\sqrt{\vec{p}^{\,2}+m_k^2}-\mu_k)/T}}
\end{equation}
and the net-baryon density follows from $n_B=\sum_k B_k n_k$. Since we consider $n_B=0$, all $\mu_k$ are set to zero in the chemical equilibrium case. 

In this work, we employ a HRG model containing states up to a mass of $2$~GeV as, for example, listed in the edition~\cite{PDG05} of the Particle Data Book. Such a list is also included in the EoS-package provided along with the work in~\cite{Huovinen09}. As evident from Fig.~\ref{Fig:1}, this choice is sufficient to describe the available lattice QCD data fairly well for temperatures below $175$~MeV, where HRG and lattice QCD results mostly overlap. In fact, the relative deviation of the HRG model from the lattice QCD data~\cite{Borsanyi11} in this overlap region, taking the error-bars in the data into account, is at most $9\%$ in $I(T)/T^4$ and $5\%$ in $p(T)/T^4$. 

Given the reasonable agreement between lattice QCD data and the HRG model, we construct an equation of state, which serves as a baseline EoS for the chemical equilibrium case: we utilize our suitable parametrization of the lattice QCD results from~\cite{Borsanyi11} at high $T$ and change the prescription to the above discussed HRG model at low $T$ around a switching temperature of $172$~MeV. Generically, such an approach can introduce discontinuities in the thermodynamic quantities. We improve this situation by employing a straightforward interpolation procedure between the two parts in the interval $165$~MeV~$\leq T\leq 180$~MeV, which ensures that the pressure and its first and second derivatives with respect to $T$ are continuous. In this way, the speed of sound remains a smooth function for all temperatures. 
A similar strategy was applied for the construction of the QCD equation of state in~\cite{Huovinen09}. 

\section{Hadron resonance gas in partial chemical equilibrium \label{sec:3}}

In heavy-ion collisions, the time scales for inelastic particle number changing processes, which are responsible for the chemical equilibration of the hadronic matter, are typically much larger than the lifetime of the hadronic stage~\cite{Teaney02}. Thus, it is more realistic to assume that the hadronic phase is not in complete chemical equilibrium. This was first discussed in~\cite{Bebie92} and then considered in numerous works, cf.~e.g.~\cite{Hirano02,Rapp02,KolbRapp03,Huovinen07,ECHO-QGP}: according to this idea, hadronic matter is formed at the hadronization temperature $T_c$ in chemical equilibrium. However, for temperatures below the chemical freeze-out temperature $T_{ch}$, where $T_{ch}\leq T_c$, the inelastic processes become suppressed, while the elastic interactions mediated by frequent strong resonance formations and decays (e.g. $\pi\pi\to\rho\to\pi\pi$, $K\pi\to K^*\to K\pi$, $p\pi\to\Delta\to p\pi$ etc) continue to occur. Consequently, the experimentally observed ratios of 
particle 
multiplicities of those species $i$, which are stable against strong decays within the lifetime of the system, are fixed at $T_{ch}$. This is to say that for $T<T_{ch}$ the corresponding effective particle numbers $\bar{N}_i=N_i+\sum_r d_{r\to i}\, N_r$ are frozen. Here, $N_i$ denotes the actual particle number of the stable hadron $i$, $N_r$ the actual particle number of resonance $r$ and $d_{r\to i}$ gives the average number of hadrons $i$ produced in the decay of resonance $r$. For example, the conserved quantity in the process $\pi\pi\to\rho\to\pi\pi$ is the effective pion number $\bar{N}_\pi\!=\!N_\pi+2N_\rho$. The above sum has to be taken over all the states (resonances) that decay into hadron $i$ within the lifetime of the hadronic stage. As their effective number is fixed at $T_{ch}$, but $T$ decreases during the expansion of matter, each stable particle species $i$ acquires an effective, $T$-dependent chemical potential $\mu_i(T)$. The chemical potentials of the resonances, instead, can be written 
as a combination $\mu_r=\sum_i d_{r\to i}\,\mu_i$ of the effective chemical potentials of the stable hadrons. The hadronic phase is, thus, in a state of partial chemical equilibrium below $T_{ch}$. 
\begin{table*}
\vspace{5mm}
\centering
 \begin{tabular}[t]{|c|c|c|c|c|c|c|c|c|}
  \hline
  \rule{0pt}{2ex}
   & \multicolumn{2}{c|}{$T_{ch}=0.145$~GeV} & \multicolumn{2}{c|}{$T_{ch}=0.150$~GeV} & \multicolumn{2}{c|}{$T_{ch}=0.155$~GeV} & \multicolumn{2}{c|}{$T_{ch}=0.160$~GeV} \\
  \hline
  \rule{0pt}{2ex}
  species & $a_i$ & $b_i$/GeV$^{-1}$ & $a_i$ & $b_i$/GeV$^{-1}$ & $a_i$ & $b_i$/GeV$^{-1}$ & $a_i$ & $b_i$/GeV$^{-1}$ \\
  \hline
  \hline
  \rule{0pt}{2.5ex}
  $\pi^0$ & 1.745 & -8.607 & 1.785 & -8.438 & 1.816 & -8.220 & 1.839 & -7.960 \\
  $\pi^+$, $\pi^-$ & 1.766 & -8.520 & 1.803 & -8.334 & 1.835 & -8.140 & 1.853 & -7.836 \\
  $K^+$, $K^-$ & 3.156 & -0.992 & 3.080 & -1.125 & 3.008 & -1.233 & 2.938 & -1.307 \\
  $K^0$, $\overline{K}^0$ & 3.191 & -1.131 & 3.114 & -1.246 & 3.044 & -1.393 & 2.973 & -1.440 \\
  $\eta$ & 3.545 & -2.127 & 3.467 & -2.296 & 3.396 & -2.465 & 3.324 & -2.538 \\
  \hline
  \rule{0pt}{2.5ex}
  $p$ & 6.104 & 1.504 & 5.893 & 1.489 & 5.694 & 1.446 & 5.507 & 1.396 \\
  $n$ & 6.113 & 1.530 & 5.899 & 1.525 & 5.701 & 1.465 & 5.513 & 1.420 \\
  $\Lambda^0$ & 6.914 & 4.951 & 6.642 & 4.730 & 6.389 & 4.466 & 6.153 & 4.202 \\
  $\Sigma^+$ & 7.393 & 2.185 & 7.145 & 1.827 & 6.915 & 1.466 & 6.700 & 1.149 \\
  $\Sigma^0$ & 7.420 & 2.160 & 7.170 & 1.806 & 6.940 & 1.453 & 6.723 & 1.145 \\
  $\Sigma^-$ & 7.460 & 2.161 & 7.211 & 1.770 & 6.977 & 1.444 & 6.760 & 1.135 \\
  $\Xi^0$ & 7.939 & 5.461 & 7.634 & 5.055 & 7.349 & 4.670 & 7.084 & 4.302 \\
  $\Xi^-$ & 7.981 & 5.551 & 7.673 & 5.149 & 7.387 & 4.748 & 7.121 & 4.373 \\
  $\Omega^-$ & 10.409 & 3.193 & 10.052 & 2.719 & 9.722 & 2.249 & 9.411 & 1.878 \\
  \hline
 \end{tabular}
 \caption[]{\label{Tab:1} Parameter-values entering Eq.~(\ref{equ:ParaOfChemPots}). With these, the effective chemical potentials $\mu_i(T)$ of the stable particle species $i$ can be described for temperatures $T<T_{ch}$. For $T\geq T_{ch}$, all $\mu_i(T)=0$. Note that anti-baryons obey the same parametrization as their respective baryons.}
\end{table*}

The freeze-out of the chemical composition of the system at $T_{ch}$ implies, in addition to the conservation of energy, momentum and of the charges $N_B$, $N_Q$ and $N_S$, also the conservation of the effective number $\bar{N}_i$ of each stable particle species $i$ below $T_{ch}$. This makes the EoS a highly-involved relation between $p$, $\epsilon$ and all charge densities. For conserved entropy, the ratio between the effective particle number density and the entropy density $\bar{n}_i/s$ is fixed at $T_{ch}$. This provides a practical tool to conserve all the $\bar{N}_i$ and to determine all the $\mu_i(T)$ for $T<T_{ch}$ from the conditions 
\begin{equation}
\label{equ:conservation}
 \frac{\bar{n}_i(T,\{\mu_{i'}(T)\})}{s(T,\{\mu_{i'}(T)\})}=\frac{\bar{n}_i(T_{ch},\{0\})}{s(T_{ch},\{0\})} \,,
\end{equation}
which imply that each $\bar{n}_i$ depends, in general, on all the effective chemical potentials $\mu_{i'}(T)$ (including $\mu_i(T)$). The knowledge of all the $\mu_i(T)$ is, apart from knowing the EoS, necessary for determining the final state hadron abundances. We note that the above conditions entail also that the particle ratios of stable hadrons are fixed at $T_{ch}$: $\bar{n}_{i1}(T,\{\mu_i\})/\bar{n}_{i2}(T,\{\mu_i\})=\bar{n}_{i1}(T_{ch},\{0\})/\bar{n}_{i2}(T_{ch},\{0\})$. 

In this work, we consider as stable particle species the mesons $\pi^0$, $\pi^+$, $\pi^-$, $K^+$, $K^-$, $K^0$, $\overline{K}^0$ and $\eta$ and the baryons $p$, $n$, $\Lambda^0$, $\Sigma^+$, $\Sigma^0$, $\Sigma^-$, $\Xi^0$, $\Xi^-$ and $\Omega^-$ as well as their respective anti-baryons, i.e.~in total $26$ different states. Correspondingly, we consider different isospin states individually. In general, this becomes important only when considering non-vanishing net-densities $n_B$, $n_Q$ and/or $n_S$. In the $n_B=0$ case studied in this work, however, particles and their corresponding anti-particles develop the same effective chemical potentials. For the chemical freeze-out temperature, we consider different values, namely $T_{ch}/$MeV$=145,\,150,\,155$ and $160$. These are within the range of the $T_c$-values determined in lattice QCD~\cite{Borsanyi09,Bazavov12}. 

\begin{figure}[t]
\centering
\vspace{5mm}
\includegraphics[width=0.45\textwidth]{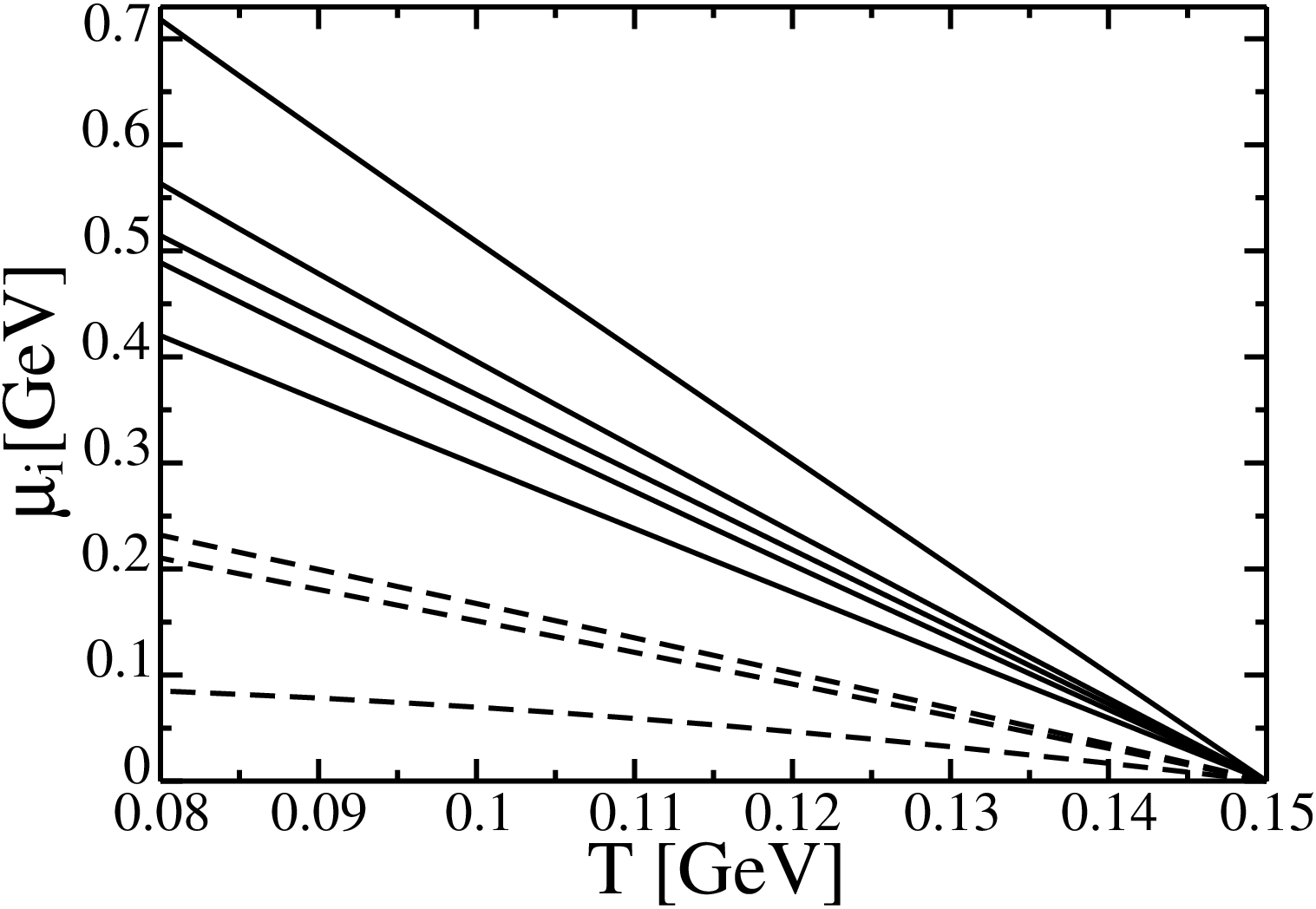} 
\caption[]{\label{Fig:2} Temperature-dependence of the effective chemical potentials for selected hadronic states, considering a chemical freeze-out temperature of $T_{ch}=150$~MeV. The solid curves depict $\mu_i(T)$ for the baryons $\Omega^-$, $\Xi^-$, $\Sigma^-$, $\Lambda^0$ and $p$ from top to bottom, while the dashed curves show $\mu_i(T)$ for the mesons $\eta$, $K^-$ and $\pi^+$ from top to bottom.}
\end{figure} 
In Fig.~\ref{Fig:2}, we exhibit the temperature-dependence of the effective chemical potentials $\mu_i(T)$ of some representative particle species as determined from Eq.~(\ref{equ:conservation}) for $T_{ch}=150$~MeV. As can be seen from Fig.~\ref{Fig:2}, the $\mu_i(T)$ increase with decreasing $T$. The $T$-dependence of $\mu_i(T)$ for species $i$ can be parametrized conveniently by the quadratic fit function 
\begin{equation}
\label{equ:ParaOfChemPots}
 \mu_i(T)=a_i(T_{ch}-T)+b_i(T_{ch}-T)^2 \,.
\end{equation}
Here, the parameters $a_i$ and $b_i$ depend on the value of $T_{ch}$. Since for a complete EoS the knowledge of all $\mu_i(T)$ is required, we summarize the corresponding parameter-values in Tab.~\ref{Tab:1}. With these, $\mu_i(T)$ is obtained in units of GeV for $T_{ch}$ and $T$ given in units of GeV. 

\begin{figure*}[ht]
\centering
\vspace{5mm}
\includegraphics[width=0.45\textwidth]{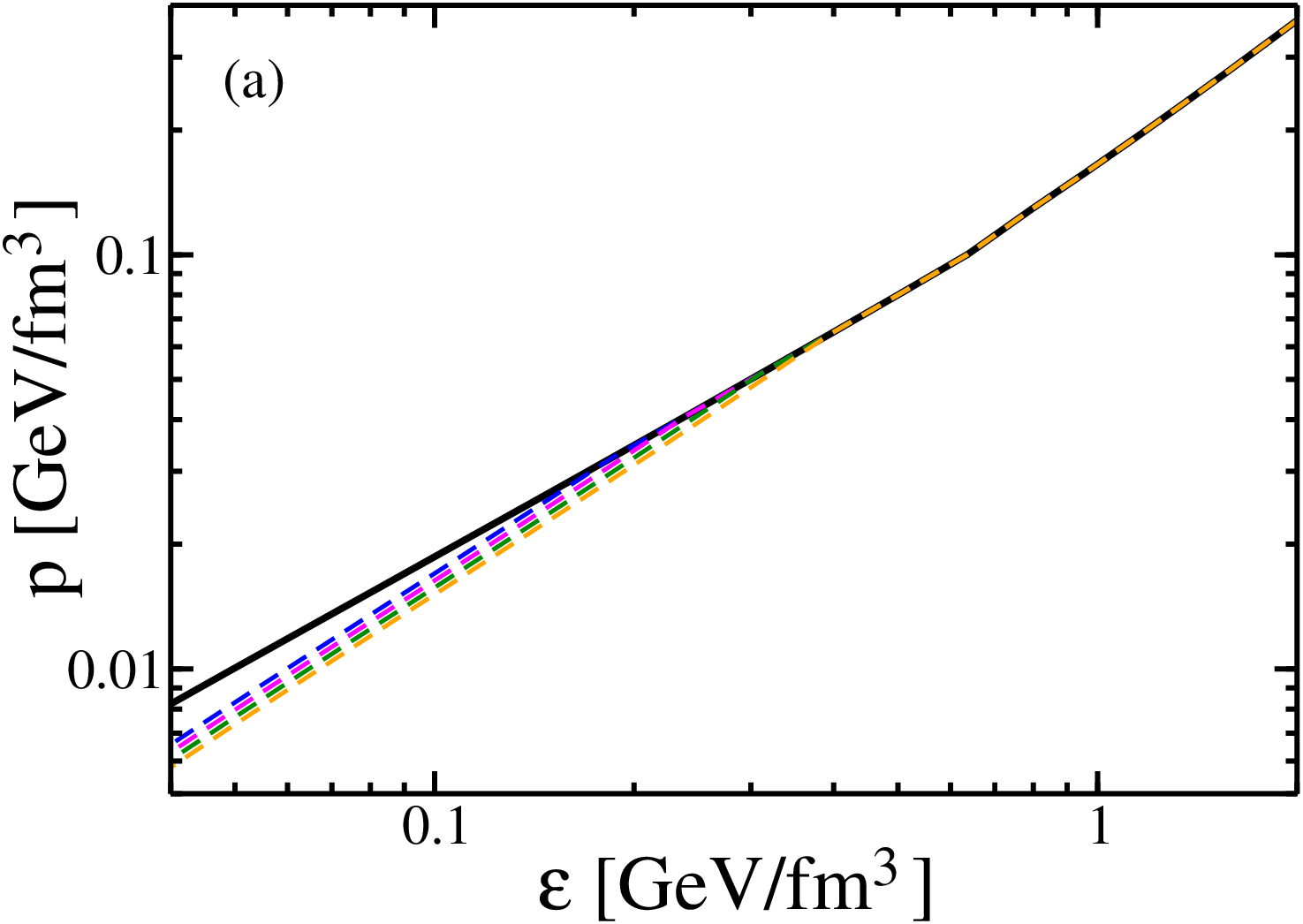} 
\hspace{3mm}
\includegraphics[width=0.45\textwidth]{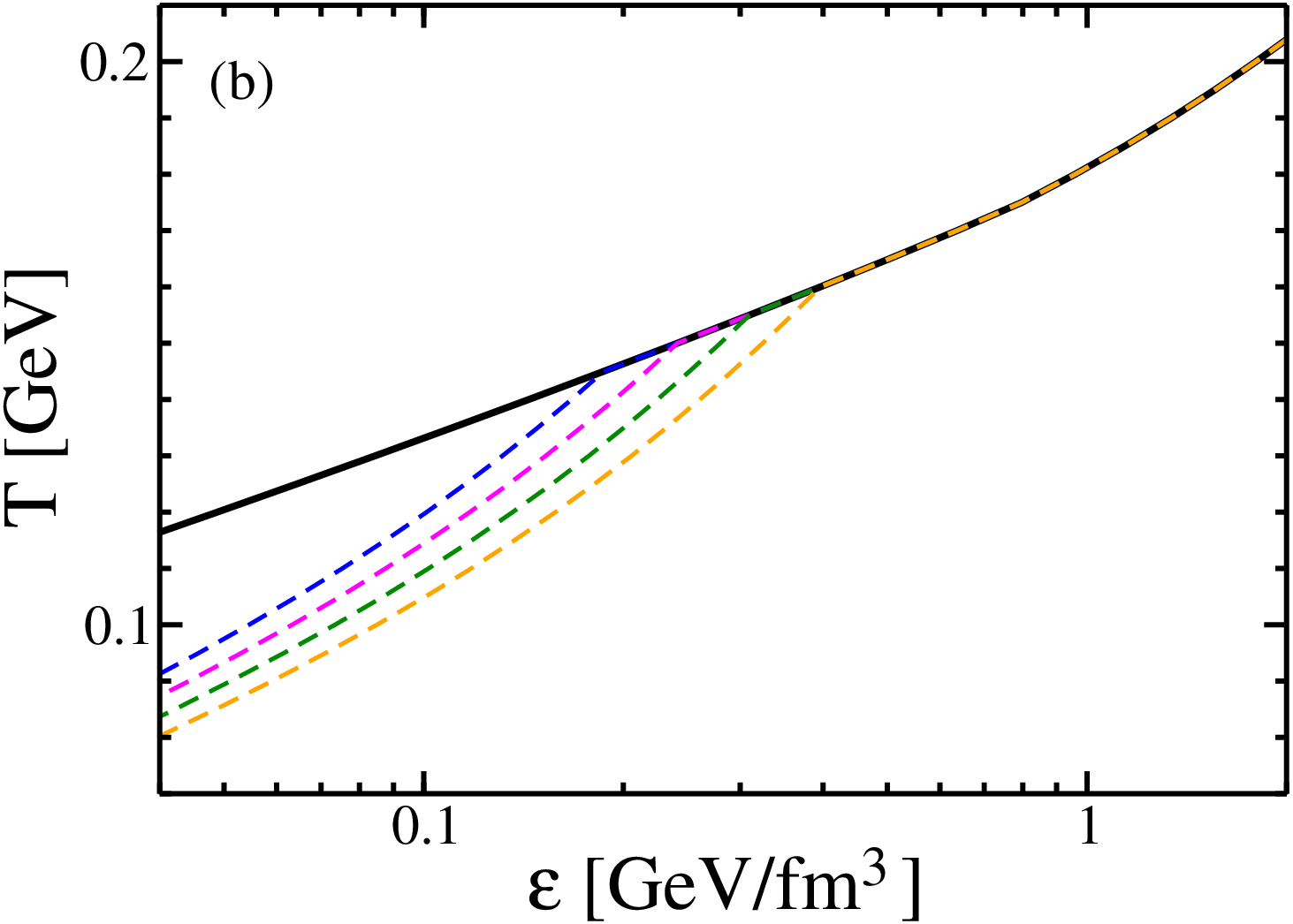} 
\caption[]{\label{Fig:3} (Color online) (a) Visualization of the different equations of state $p(\epsilon)$ for $n_B=0$ zoomed into the regions in $\epsilon$, in which the confinement transition and the chemical freeze-out occur. The solid curve depicts the EoS with a HRG in chemical equilibrium in the hadronic phase. The dashed curves show the equations of state including partial chemical equilibrium below $\epsilon_{ch}$. The value of $\epsilon_{ch}$ depends on the value of the freeze-out temperature $T_{ch}$. We consider $T_{ch}/$MeV$\,=145,\,150,\,155$ and $160$ (from top to bottom in the figure, respectively). (b) Connection between temperature and energy density $T(\epsilon)$ for the equations of state with chemical equilibrium (solid curve) and with partial chemical equilibrium (dashed curves) in the hadronic phase (labeling as in panel (a)). Suitable parametrizations of these results as functions of $\epsilon$ are provided in the Eqs.~(\ref{equ:ParamPressure})~-~(\ref{equ:ParamTemperature}) together 
with Tab.~\ref{Tab:2} and in the Eqs.~(\ref{equ:ParamPressureFO}) and~(\ref{equ:ParamTemperatureFO}) together with Tab.~\ref{Tab:3}.}
\end{figure*} 
These parametrizations provide excellent fits for all $\mu_i(T)$ in the temperature range $70$~MeV~$\leq T\leq T_{ch}$ with a maximal $\chi^2=9\cdot 10^{-6}$. We note that overall, for the large $T$-range explored in a hydrodynamic simulation, cubic fit functions for the $\mu_i(T)$ yield more accurate descriptions of the full numerical results obtained from Eq.~(\ref{equ:conservation}) than the quadratic functions, in particular for small $T$. In the interesting interval $70$~MeV~$\leq T\leq T_{ch}$, however, the quadratic ansatz Eq.~(\ref{equ:ParaOfChemPots}) provides fits, which are comparable in accuracy with the cubic-fits for the baryons and anti-baryons, while they are even slightly better for the mesons. 

\section{Discussion and Conclusions \label{sec:4}}

We obtain various equations of state by combining our parametrization of the lattice QCD data~\cite{Borsanyi11} as a function of $T$ with the HRG model either in chemical equilibrium or in partial chemical equilibrium in the hadronic phase with various $T_{ch}$-values in the latter case. For the use in a hydrodynamic simulation, however, the EoS is usually given in the form $p(\epsilon,n_B)$, i.e.~as a function of $\epsilon$ and $n_B$, together with the results for the effective chemical potentials $\mu_i$ required for determining the particle abundances. In Fig.~\ref{Fig:3}, we show our results for the different equations of state $p(\epsilon)$ supplemented by the corresponding $T(\epsilon)$ for $n_B=0$. We concentrate in Fig.~\ref{Fig:3} on a visualization of the energy density regions, in which the confinement transition and the chemical freeze-out take place. 

As it is evident from Fig.~\ref{Fig:3} panel (a), differences in $p(\epsilon)$ between chemical equilibrium (solid curve) and partial chemical equilibrium (dashed curves) in the hadronic phase are small. The $\epsilon$-dependence of $T$, instead, is visibly influenced for $\epsilon<\epsilon_{ch}$ by the chemical freeze-out (cf. panel (b) in Fig.~\ref{Fig:3}), where the value of $\epsilon_{ch}$ depends on $T_{ch}$. 

Our results are collected in tabulated form and made available along with this publication~\cite{webpage}. Moreover, for practical convenience we also provide parametrizations as a function of $\epsilon$ of these numerical results, similar to Ref.~\cite{Shen10}. In the chemical equilibrium case, the relevant thermodynamic quantities can be parametrized in the following way: 
\begin{eqnarray}
\nonumber
 p(\epsilon) & = & a_0\epsilon +\frac{a_1}{(a_2+1)}\,\epsilon^{a_2+1}+\frac{a_3}{a_4}\exp\left[a_4\epsilon\right] \\
\label{equ:ParamPressure}
 & & -\frac{a_5}{(-a_7)^{a_6+1}}\,\Gamma(a_6+1,-a_7\epsilon) + a_8 \,, \\
\nonumber
 s^{4/3}(\epsilon) & = & a_0+a_1\epsilon^{a_2} +a_3\exp\left[a_4\epsilon\right] \\
\label{equ:ParamEntropy} 
 & & + a_5\epsilon^{a_6}\exp\left[a_7\epsilon\right]
\end{eqnarray}
and 
\begin{equation}
\label{equ:ParamTemperature}
 T(\epsilon) = \frac{\epsilon+p(\epsilon)}{s(\epsilon)} \equiv \frac{1}{ds(\epsilon)/d\epsilon} \,.
\end{equation}
Here, $\Gamma(s,x)=\int_x^\infty t^{s-1} \exp[-t]\,dt$ denotes the upper incomplete $\Gamma$-function. These ansatz-functions can provide excellent descriptions of our numerical EoS-results with proper choices for the entering parameters. We stress, that the parameters $a_i$ in $p(\epsilon)$ and in $s^{4/3}(\epsilon)$ in Eqs.~(\ref{equ:ParamPressure}) and~(\ref{equ:ParamEntropy}) are not meant to be the same: we use the same symbols only for practical purposes. 

\begin{table*}
\vspace{5mm}
\centering
 \begin{tabular}[t]{|c|c|c|c|c|c|c|c|c|c|c|}
  \hline
  \rule{0pt}{2ex}
  quantity & $\epsilon$-region & $a_0$ & $a_1$ & $a_2$ & $a_3$ & $a_4$ & $a_5$ & $a_6$ & $a_7$ & $a_8$ \\
  \hline
  \hline
  \rule{0pt}{2.5ex}
  $p$ & $\epsilon<\epsilon_0$            & 0.275255 & 2524790 & 2711.84 & -0.275255 & -274.84 & 0.487526 & 0.0956908 & -388.771 & -0.000326 \\
      & $\epsilon_0<\epsilon<\epsilon_1$            & 0.843569 & -60.3954 & 3.203 & -0.601971 & 2.06599 & -739.605 & 2.15326 & -122.409 & 0.2909065 \\
      & $\epsilon_1<\epsilon<\epsilon_2$ & 4.7406 & -4.1849 & 0.1807 & -5.4941 & -1.8539 & 4.3735 & 0.1003 & -2.3275 & -1.321345 \\
      & $\epsilon_2<\epsilon$            & $1/3$ & -0.1310034 & -0.4179 & -0.0230894 & -0.2797 & -0.0774 & 5.7231 & -3.3064 & -0.018477 \\
  \hline
  \rule{0pt}{2.5ex}
  $s^{4/3}$ & $\epsilon<\epsilon_0$            & 2.12885 & 12.50217 & 1.07208 & -2.12885 & 1.0032 & 0.0084625 & 1.42094 & 1345.43 &  \\
            & $\epsilon_0<\epsilon<\epsilon_1$            & -0.000165 & 9.1583717 & 1.0786 & 0. & 1. & 0.5649 & 1.0959 & 9.9955 &  \\
            & $\epsilon_1<\epsilon<\epsilon_2$ & -0.0003645 & 5.763101 & 1.3863 & -0.0000745 & 0.3105 & 6.7934 & 1.0337 & -0.0976 &  \\
            & $\epsilon_2<\epsilon<\epsilon_3$ & -0.655216 & 18.36345 & 0.9912019 & 2.02343 & 0.00355427 & -7.78303 & 0.5725142 & 0.0039527 &  \\
            & $\epsilon_3<\epsilon<\epsilon_4$ & 1.49791 & 14.83324 & 1.02947504 & 3.244652 & -0.0372865 & -7.96072 & 0.257059 & -0.056146 &  \\
            & $\epsilon_4<\epsilon<\epsilon_5$ & -19.025 & 16.25163 & 1.012862 & 33.989528 & -0.00763319 & -18.69299 & 0.179563 & -0.009929 &  \\
            & $\epsilon_5<\epsilon$            & -33.07911 & 16.978858 & 1.00512 & 0. & 1. & 0. & 1. & 1. &  \\
  \hline
 \end{tabular}
 \caption[]{\label{Tab:2} Summary of the parameter-values entering Eqs.~(\ref{equ:ParamPressure}) and~(\ref{equ:ParamEntropy}) for $p(\epsilon)$ and $s^{4/3}(\epsilon)$, respectively, providing practical parametrizations of our numerical results for the case of chemical equilibrium in the hadronic phase. The fits are optimized in different $\epsilon$-regimes, where $\epsilon_0=0.001538$~GeV$/$fm$^3$,  $\epsilon_1=0.032084$~GeV$/$fm$^3$, $\epsilon_2=0.567420$~GeV$/$fm$^3$, $\epsilon_3=1.2$~GeV$/$fm$^3$, $\epsilon_4=9.9$~GeV$/$fm$^3$ and $\epsilon_5=100$~GeV$/$fm$^3$. The parameters are given in such units that for $\epsilon$ in GeV$/$fm$^3$ one finds $p$ in units of GeV$/$fm$^3$ and $s$ in units of $1/$fm$^3$.}
\end{table*}
It turns out that, for an accurate description of the thermodynamic quantities, it becomes mandatory to split the parametrizations into different regions in $\epsilon$ and to fit the parameters for each $\epsilon$-region individually. We define as the splitting-points $\epsilon_0=0.001538$~GeV$/$fm$^3$, $\epsilon_1=0.032084$~GeV$/$fm$^3$, $\epsilon_2=0.567420$~GeV$/$fm$^3$, $\epsilon_3=1.2$~GeV$/$fm$^3$, $\epsilon_4=9.9$~GeV$/$fm$^3$ and $\epsilon_5=100$~GeV$/$fm$^3$. The points $\epsilon_3$, $\epsilon_4$ and $\epsilon_5$ are only of relevance for the parametrization of 
$s^{4/3}(\epsilon)$ in Eq.~(\ref{equ:ParamEntropy}) and, therefore, influence $T(\epsilon)$, but play no role for the parametrization of $p(\epsilon)$. The parameter-values entering $p(\epsilon)$ and $s^{4/3}(\epsilon)$ in the different $\epsilon$-regimes are summarized in Tab.~\ref{Tab:2}. With these, one obtains $p$ in units of GeV$/$fm$^3$, $s$ in units of $1/$fm$^3$ and $T$ from Eq.~(\ref{equ:ParamTemperature}) in units of GeV for $\epsilon$ given in units of GeV$/$fm$^3$. 
The high precision in the provided parametrizations is motivated by our goal to maintain thermodynamic consistency and continuity in the second derivatives at the splitting-points $\epsilon_i$ up to a high numerical accuracy. 

The squared speed of sound $c_s^2(\epsilon)$ as a function of $\epsilon$ can be determined from $p(\epsilon)$ given in Eq.~(\ref{equ:ParamPressure}) as 
\begin{equation}
\label{equ:ParamSpeed}
 c_s^2(\epsilon) = a_0+a_1\epsilon^{a_2} +a_3\exp\left[a_4\epsilon\right] + a_5\epsilon^{a_6}\exp\left[a_7\epsilon\right] \,.
\end{equation}
By employing the parameter-values for $p(\epsilon)$ from Tab.~\ref{Tab:2}, we find an excellent agreement between Eq.~(\ref{equ:ParamSpeed}) and the $c_s^2$-result obtained by numerically differentiating our tabulated $p(\epsilon)$-results within the temperature interval $30$~MeV~$\leq T\leq 300$~MeV. Outside of this range, the quantitative agreement is still good, where $c_s^2$ exhibits the same qualitative behavior as our numerical results with asymptotics $c_s^2(\epsilon)\to 0$ for $\epsilon\to 0$ and $c_s^2(\epsilon)\to 1/3$ for $\epsilon\to\infty$. 

\begin{figure}[t]
\centering
\vspace{5mm}
\includegraphics[width=0.45\textwidth]{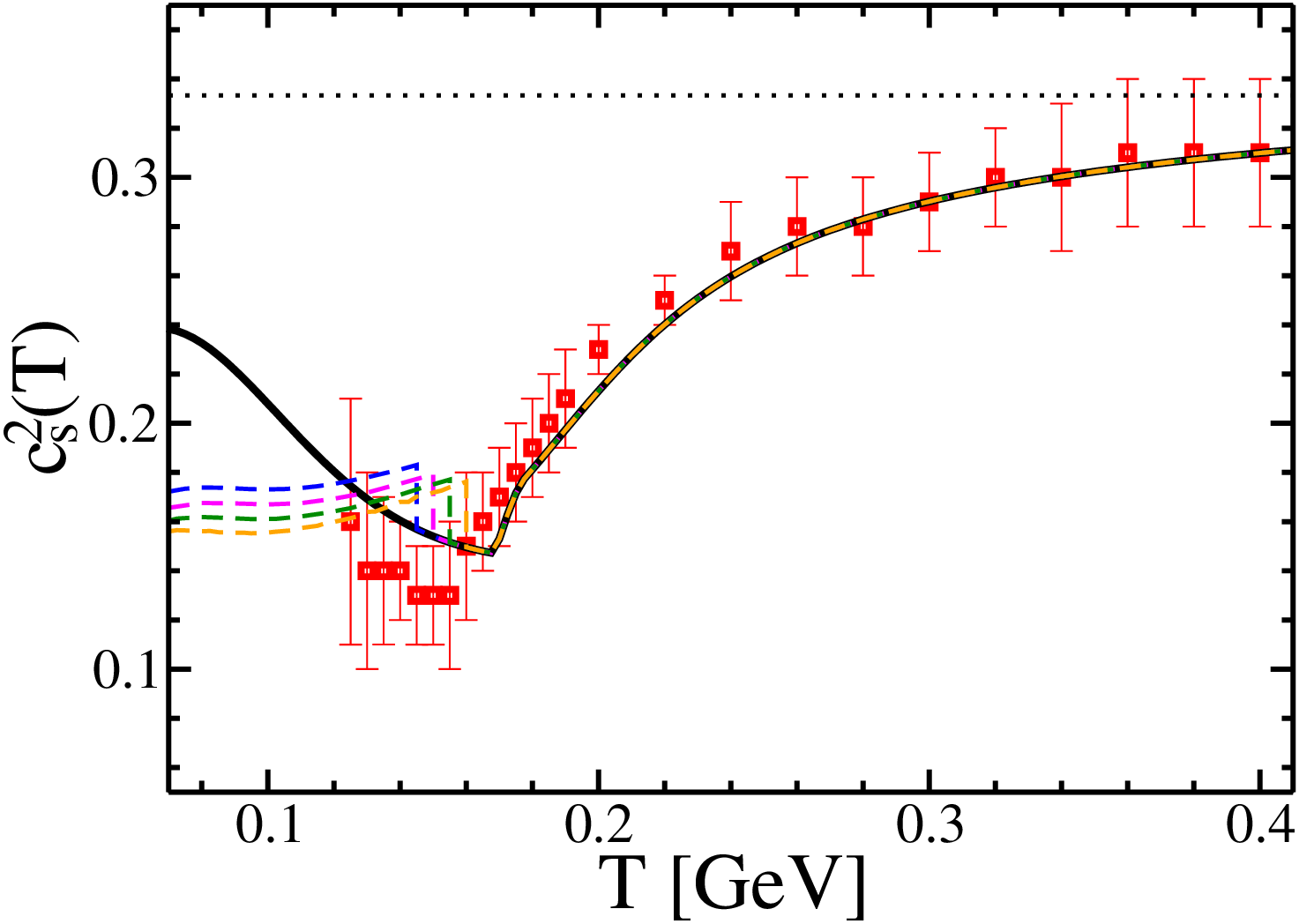} 
\caption[]{\label{Fig:4} (Color online) Temperature-dependence of the squared speed of sound $c_s^2(T)$. The solid curve shows the result obtained from a numerical differentiation of our tabulated $p(\epsilon)$-results for the EoS, in which the HRG is in chemical equilibrium. For comparison, the symbols depict available equilibrium lattice QCD data from~\cite{Borsanyi10}. The dashed curves highlight $c_s^2(T)$ when instead partial chemical equilibrium is assumed in the hadronic phase. We consider $T_{ch}/$MeV$\,=145,\,150,\,155$ and $160$ (from top to bottom, respectively).}
\end{figure}

The temperature-dependence of $c_s^2$ obtained by numerical differentiation is shown in Fig.~\ref{Fig:4} (solid curve) and confronted with the lattice QCD results available from the WB-collaboration~\cite{Borsanyi10}. Our curve agrees with the lattice QCD data within error-bars: we also find a rather large $c_s^2(T)$ in the confinement transition region. This indicates that our EoS is rather stiff compared to some previously considered equations of state, as e.g.~in~\cite{Song08}, but comparable in stiffness with the equation of state presented in~\cite{Huovinen09}. 

\begin{table*}
\vspace{5mm}
\centering
 \begin{tabular}[t]{|c|c|c|c|c|c|c|c|}
  \hline
  \rule{0pt}{2ex}
  $T_{ch}/$GeV & quantity & $b_0$ & $b_1$ & $b_2$ & $b_3$ & $b_4$ & $b_5$ \\
  \hline
  \hline
  \rule{0pt}{2.5ex}
  0.145 & $p$       &  & 0.20421265 & 1.2147 & -0.006941 & -10.9535 & \\
        & $s^{4/3}$ & -5.727246 & 17.2995 & 1.0833 &  &  & \\
        & $T$       & -0.0570889 & 0.5799 & 0.1412 & -0.3744 & 0.0923 & -0.3201 \\
  \hline
  \rule{0pt}{2.5ex}
  0.150 & $p$       &  & 0.19620877 & 1.2200 & -0.007064 & -10.4533 & \\
        & $s^{4/3}$ & -4.855376 & 16.5519 & 1.0928 &  &  & \\
        & $T$       & -0.838232 & 1.9880 & 0.1842 & -1.7244 & 0.1689 & -0.5845 \\
  \hline
  \rule{0pt}{2.5ex}
  0.155 & $p$       &  & 0.18633308 & 1.2194 & -0.00730055 & -9.6563 & \\
        & $s^{4/3}$ & -4.4274 & 16.2165 & 1.0998 &  &  & \\
        & $T$       & -1.4628554 & 2.5136 & 0.2741 & -2.1268 & 0.2575 & -0.8329 \\
  \hline
  \rule{0pt}{2.5ex}
  0.160 & $p$       &  & 0.17665061 & 1.2121 & -0.00726805 & -9.0082 & \\
        & $s^{4/3}$ & -4.1890264 & 16.0463 & 1.1055 &  &  & \\
        & $T$       & -1.2085036 & 2.5073 & 0.2522 & -2.1677 & 0.2371 & -0.6475 \\
  \hline
 \end{tabular}
 \caption[]{\label{Tab:3} Summary of the parameter-values entering Eqs.~(\ref{equ:ParamPressureFO}), (\ref{equ:ParamEntropyFO}) and~(\ref{equ:ParamTemperatureFO}) for $p(\epsilon)$, $s^{4/3}(\epsilon)$ and $T(\epsilon)$, respectively, providing practical parametrizations of our numerical results for $\epsilon<\epsilon_{ch}$ when partial chemical equilibrium is considered in the hadronic phase. For $\epsilon>\epsilon_{ch}$, the thermodynamic quantities are given by Eqs.~(\ref{equ:ParamPressure})~-~(\ref{equ:ParamTemperature}) together with Tab.~\ref{Tab:2}. The optimized fit-parameters depend on the value of $\epsilon_{ch}$, which varies with $T_{ch}$, where $\epsilon_{ch}=0.18675523$~GeV$/$fm$^3$ for $T_{ch}=0.145$~GeV, $\epsilon_{ch}=0.24117503$~GeV$/$fm$^3$ for $T_{ch}=0.150$~GeV, $\epsilon_{ch}=0.30993163$~GeV$/$fm$^3$ for $T_{ch}=0.155$~GeV and $\epsilon_{ch}=0.39623763$~GeV$/$fm$^3$ for $T_{ch}=0.160$~GeV. The parameters are given in units, such that for $\epsilon$ in 
GeV$/$fm$^3$ one finds $p$ in units of GeV$/
$fm$^3$, $s$ in units of $1/$fm$^3$ and $T$ in units of GeV.}
\end{table*}
When including partial chemical equilibrium into the EoS, the parametrizations discussed above have to be modified only for $\epsilon<\epsilon_{ch}$. The different values of $\epsilon_{ch}$, depending on the considered value for the chemical freeze-out temperature, are listed in the caption of Tab.~\ref{Tab:3}. For $\epsilon<\epsilon_{ch}$, we modify our parametrizations to 
\begin{eqnarray}
\label{equ:ParamPressureFO}
 p(\epsilon) & = & b_1\epsilon^{b_2} + b_3\left(\exp\left[b_4\epsilon\right]-1\right) \,, \\
\label{equ:ParamEntropyFO} 
 s^{4/3}(\epsilon) & = & b_0\epsilon + b_1\epsilon^{b_2} 
\end{eqnarray}
and 
\begin{equation}
\label{equ:ParamTemperatureFO}
 T(\epsilon) = b_0\epsilon + b_1\epsilon^{b_2} + b_3\epsilon^{b_4}\exp\left[b_5\epsilon\right] \,.
\end{equation}
Correspondingly, the squared speed of sound follows now from Eq.~(\ref{equ:ParamPressureFO}) as 
\begin{equation}
\label{equ:ParamSpeedFO}
 c_s^2(\epsilon) = b_1b_2\epsilon^{b_2-1}+b_3b_4\exp\left[b_4\epsilon\right] \,.
\end{equation}
We stress that the parameters $b_i$ entering $p(\epsilon)$, $s^{4/3}(\epsilon)$ and $T(\epsilon)$ in Eqs.~(\ref{equ:ParamPressureFO})~-~(\ref{equ:ParamTemperatureFO}) are also here not meant to be the same. 

By fitting the parametrizations in Eqs.~(\ref{equ:ParamPressureFO})~-~(\ref{equ:ParamTemperatureFO}) to our tabulated, numerical results for $\epsilon<\epsilon_{ch}$, we find quite accurate descriptions of the thermodynamic quantities for the parameter-values summarized in Tab.~\ref{Tab:3}. Again, the high precision in the parameters is given in order to maintain consistency in our approach at $\epsilon_{ch}$ with a high-level of accuracy. With these parameters, $p$, $s$ and $T$ are obtained in units of GeV$/$fm$^3$, $1/$fm$^3$ and GeV, respectively, for $\epsilon$ given in units of GeV$/$fm$^3$. 
Moreover, the above parametrizations satisfy the physical conditions $T(\epsilon)\to 0$, $p(\epsilon)\to 0$ and $s(\epsilon)\to 0$ for $\epsilon\to 0$. The qualitative behavior of the squared speed of sound is also nicely reproduced with, however, different asymptotics (in fact one obtains positive $c_s^2(\epsilon\to 0)=b_3b_4<1/3$, while our numerical results tend toward $0$). 

The temperature-dependence of $c_s^2$ as determined from a numerical differentiation of our tabulated results is shown in Fig.~\ref{Fig:4} (dashed curves). One observes a discontinuity in $c_s^2(T)$ at $T=T_{ch}$, which is characteristic for the chemical freeze-out. Evidently, as expected the behavior of $c_s^2(T)$ in the non-equilibrium situation is different from the trend seen in equilibrium lattice QCD thermodynamics. 

In summary, we constructed QCD equations of state for vanishing net-baryon density based at high $T$ on recent continuum-extrapolated lattice QCD results in the physical quark mass limit~\cite{Borsanyi11}, which were continuously combined with a HRG model at low $T$. The latter was considered to be either in chemical equilibrium or in partial chemical equilibrium. In the chemical equilibrium case, our baseline EoS in terms of the squared speed of sound shows only minor deviations of at most a few percent from the EoS presented in~\cite{Huovinen09} such that, presumably, significant differences in the standard observables studied in hydrodynamic simulations are not to be expected from this EoS. Nevertheless, the focus of our work lay on the inclusion of partial chemical equilibrium for a more accurate description of the experimental situation, 
where we studied different values for the chemical freeze-out temperature $T_{ch}$ within the range presently expected from first-principle approaches~\cite{Borsanyi13}. In view of the non-negligible differences in the temperature and chemical potential evolution of the system for different $T_{ch}$-values our work, thus, allows one to study the possible influence of deviations in $T_{ch}$ on the particle spectra in more detail compared to the work presented in~\cite{Shen10,Huovinen09}. 

Our results, being available in a tabulated form~\cite{webpage}, can be directly applied in the hydrodynamic modeling of high-energy heavy-ion collisions at the LHC and at RHIC for top beam energies. For convenience, we also provided practical parametrizations of our results, in particular, for the effective chemical potentials $\mu_i(T)$ of the stable hadrons in the partial chemical equilibrium case and for the temperature. Their knowledge is crucial for a determination of final state hadron abundances and spectra. 

We have restricted ourselves to the $n_B=0$ case in this work. In general, however, our approach allows for respecting the conservation of finite values for $n_B/s$ and $n_Q/s$ (while $n_S/s=0$) as relevant for heavy-ion collisions. Corresponding results for non-zero (although not too large) values of the associated chemical potentials will be reported in a forthcoming publication. 

\section*{Acknowledgements}
We acknowledge valuable discussions with P.~Huovinen and U.~Heinz. 
The work of C. Ratti and M. Bluhm is supported 
by funds provided by the Italian Ministry of Education, Universities and Research under the
Firb Research Grant RBFR0814TT.

\end{document}